\documentclass[a4paper,11pt]{article}
\usepackage{jheppub}
\usepackage{lineno}
\usepackage{orcidlink}
\usepackage{etoolbox}
\usepackage{booktabs}
\usepackage{amsmath}
\allowdisplaybreaks[4]

\title{\boldmath Holographic SU(3) color superconductivity from color-octet condensation in AdS$_4$}

\author[a,b]{Xin Zhao\orcidlink{0009-0004-0523-6191},}
\author[a,\ast]{Zhang-Yu Nie\orcidlink{0000-0001-7064-247X},\note[$\ast$]{Corresponding author.}}
\author[b]{Ya-Peng Hu\orcidlink{0000-0002-3394-2614}}

\affiliation[a]{Center for Gravitation and Astrophysics, Faculty of Science, Kunming University of Science and Technology, Kunming 650500, China}
\affiliation[b]{Center for the Cross-disciplinary Research of Space Science and Quantum-technologies (CROSS-Q), College of Physics, Nanjing University of Aeronautics and Astronautics, 29 Jiangjun Road, Nanjing City, Jiangsu Province 211106, China}

\emailAdd{zhaox923@nuaa.edu.cn}
\emailAdd{niezy@kust.edu.cn}
\emailAdd{huyp@nuaa.edu.cn}

\abstract{We present a holographic model of $SU(3)$ color superconductivity in asymptotically AdS$_4$ in which scalar and vector color-octet condensates form at finite baryon chemical potential. At $\mu_b=0$, a channel scan within consistent truncations shows that the favored sector reproduces the known $SU(2)$ s+p competition. At finite $\mu_b$, the s-wave order generalizes to an $S_1+iS_2$ structure. In the probe limit, the $S_+$ critical temperature grows linearly with $\mu_b/\mu_3$, while the p-wave critical temperature is independent of $\mu_b$. Once backreaction is included, the p-wave critical temperature decreases with $\mu_b$ and the p-wave phase disappears beyond a critical value of $\mu_b$. At high $\mu_b$, the balanced $S_+$ phase (with $S_1=S_2$) is the dominant color-superconducting phase, while the $S_-$ condensate, which signals the imbalance between $S_1$ and $S_2$, appears only at low temperature and low baryon density. Our results provide a holographic laboratory for color-octet condensation and non-Abelian color dynamics in dense matter such as neutron-star interiors.}

\begin{document}
\maketitle
\flushbottom
	
\section{Introduction}
\label{sec:intro}
Quantum chromodynamics (QCD) exhibits a rich and intricate phase structure governed by its non-Abelian $SU(3)$ gauge symmetry. In the high-density, low-temperature regime, the formation of a diquark condensate $\langle qq \rangle$ is expected to spontaneously break the color $SU(3)$ symmetry, leading to the phenomenon known as color superconductivity (CSC)~\cite{Alford:1997zt}. This exotic state of matter is predicted to exist in the interiors of neutron stars~\cite{Geissel:2025vnp}, and understanding its properties is a key objective in modern nuclear and particle physics (for comprehensive reviews, see Ref.~\cite{Alford:2007xm}). Direct experimental access to this region is challenging, and conventional lattice QCD techniques struggle at large chemical potentials, making gauge/gravity duality indispensable for exploring its properties.

Over the past decades, gauge/gravity duality has been successfully applied to model the superconductor phase transitions accompanied by the spontaneous breaking of $U(1)$ gauge symmetry~\cite{Gubser:2008px,Hartnoll:2008vx, Hartnoll:2008kx, Gubser:2008wv, Herzog:2008he, Cai:2013aca, Zhao:2022jvs} and to study the far-from-equilibrium dynamics~\cite{Bhaseen:2012gg, Adams:2012pj, Li:2020ayr, Xia:2019eje, Zeng:2019yhi, Yang:2019ibe, Baggioli:2019mck, Baggioli:2021tzr, Li:2019swh, Guo:2018mip, Lan:2023gyc, Yang:2023dvk, Zhao:2023ffs}. It has also been actively developed to probe the strongly coupled QCD problems~\cite{Erlich:2005qh,Karch:2006pv,Gursoy:2008bu,Cai:2022omk} including CSC~\cite{Chen:2009kx, Basu:2011yg,Faedo:2018fjw,BitaghsirFadafan:2018iqr,Ghoroku:2019trx, Nam:2021ufk,Preau:2025ubr,CruzRojas:2025fzs}. However, most holographic CSC constructions adopt an Abelian Higgs mechanism, effectively reducing the non-Abelian color group to a $U(1)$ symmetry. While tractable, this approach fails to capture the essential non-Abelian Higgsing of $SU(3)_c$ by a diquark condensate, which leads to characteristic breaking patterns such as the two-flavor superconducting (2SC) and color-flavor-locked (CFL) phases~\cite{Chen:2009kx}. This limitation obscures genuinely non-Abelian phenomena and motivates a refined holographic treatment. A more faithful holographic description should therefore retain the non-Abelian color dynamics, which has not been studied in this bottom-up octet framework.

To address this gap, we construct a bottom-up holographic model in asymptotically AdS$_4$ that couples a scalar octet to an $SU(3)$ color gauge field and a $U(1)_b$ field dual to the baryon-conserved current. The scalar octet is dual to a color-octet operator, which is well motivated in beyond-the-Standard-Model (BSM) physics~\cite{Manohar:2006ga}. Its condensation provides a complementary non-Abelian channel to the QCD diquark CSC, rather than a literal diquark description. The 2SC and CFL phases require colored scalars in the $\bar{\mathbf 3}$ or $\mathbf 6$ representation and are left for future work. For the favored channel identified in Sec.~\ref{sec2}, the model reproduces the known $SU(2)$ s+p competition at zero baryon chemical potential $\mu_b$. At finite $\mu_b$, the s-wave order generalizes to a two-component $S_1+iS_2$ structure, and we obtain the $\mu_b$–$T$ phase diagram. This setup provides a bottom-up laboratory for non-Abelian color dynamics and the associated response of the colored sector, enabling systematic comparisons with QCD-motivated holographic approaches to CSC~\cite{Basu:2011yg, BitaghsirFadafan:2020otb, CruzRojas:2025fzs}.

This paper is organized as follows. In Sec.~\ref{sec2}, we establish the holographic $SU(3)_c \times U(1)_b$ framework and systematically analyze the thermodynamic ground state in the absence of baryon density ($\mu_b=0$), performing an exhaustive stability screening across all channels to identify the favored condensation ansatz. In Sec.~\ref{sec3}, we extend this favored channel to finite baryon chemical potential, formulate the complete set of coupled nonlinear equations of motion incorporating gravitational backreaction, elucidate the decoupling transformation along with the resulting effective chemical potentials, and outline typical condensation behaviors. In Sec.~\ref{sec4}, we investigate the global phase diagrams in the $\mu_b$–$T$ plane, exploring the competition between the $S_1+iS_2$ and p-wave orders, clarifying the analytical scaling laws in the probe limit and studying the nontrivial effects induced by gravitational backreaction. Finally, in Sec.~\ref{sec5}, we discuss our results and outline future perspectives.

\section{The Holographic Setup and Channel Selection}\label{sec2}
In this section, we establish the $SU(3) \times U(1)$ holographic framework for color-octet condensation and, working in the probe limit at zero baryon chemical potential, perform a channel selection to identify the favored condensation channel.

\subsection{The holographic model for $SU(3) \times U(1)$}\label{sec2.1}
To model color-octet condensation at finite baryon chemical potential, we introduce on the gravity side a scalar octet charged under both $U(1)_b$ and $SU(3)_c$ gauge fields. The total action is given by
\begin{align}
	S=&\,S_{G}+S_{M}~,\\ 
    S_G=&\,\frac{1}{2\kappa_g ^2}\int d^{4}x\sqrt{-g}(R-2\Lambda)~,\\
	S_M=&\,\frac{1}{g_{c}^{2}}\int d^{4}x\sqrt{-g}\Big(-\frac{1}{4}F_{\mu\nu}^a F^{a\mu\nu}-\frac{1}{4}\tilde{F}_{\mu\nu} \tilde{F}^{\mu\nu}-\tilde{D}_{\mu}\Psi^{a}(\tilde{D}^{\mu}\Psi^{a})^*-m^{2}|\Psi^{a}|^2 \Big)~.\label{S_SU3U1}
\end{align}
$\Lambda=-3/L^2$ is the negative cosmological constant, with $L$  the AdS radius. $F^{a}_{\mu\nu}=\nabla_{\mu}A_{\nu}^{a}-\nabla_{\nu}A_{\mu}^{a}+f^{abc} A_{\mu}^{b}A_{\nu}^{c}$ is the $SU(3)$ field strength, where $f^{abc}$ are the structure constants of $SU(3)_c$. $\tilde{F}_{\mu\nu}=\nabla_{\mu}\tilde{A}_{\nu}-\nabla_{\nu}\tilde{A}_{\mu}$ is the $U(1)$ field strength. The scalar octet is in the adjoint representation of $SU(3)_c$, and $g_c$ is the Yang-Mills coupling. The covariant derivative of the scalar field is
\begin{align}
\tilde{D}_{\mu}\Psi^{a}=\nabla_{\mu}\Psi^{a}+f^{abc}A_{\mu}^{b}\Psi^{c}-\frac{2}{3}i\tilde{A}_{\mu}\Psi^{a}~.
\end{align}

Varying the full action yields the following equations of motion for the matter fields:
\begin{align}
        g^{\mu \nu} \tilde{D}_{\mu} \tilde{D}_{\nu} \Psi^{a}- m^2 \Psi^{a}&= 0~,\label{EoM_psi_vec}\\
	    g^{\rho \nu} D_{\rho}F_{\nu\mu}^{a}-f^{abc}(\Psi^{b})^*D_{\mu}\Psi^{c}-f^{abc}\Psi^{b}(D_{\mu}\Psi^{c})^*&=0~,\label{EoM_At_SU3_vec}\\
        g^{\rho \nu} \nabla_{\rho} \tilde{F}_{\mu \nu}+\frac{8}{9} \tilde{A}_{\mu} |\Psi^{a}|^2+\frac{2}{3}i\Big((\Psi^{a})^*D_{\mu}\Psi^{a}-\Psi^{a}(D_{\mu}\Psi^{a})^*\Big)&=0~.\label{EoM_At_U1_vec}
\end{align}
Here, $D_\mu T_{\alpha \beta\dots }^{a}=\nabla_{\mu}T_{\alpha \beta\dots }^{a}+f^{abc}A_{\mu}^{b}T_{\alpha \beta\dots }^{c}$ is the covariant derivative operator related to the $SU(3)$ gauge fields in the curved spacetime. 

The equations for the gravitational fields are:
\begin{align}
        R_{\mu \nu}-\frac{1}{2}(R-2\Lambda)g_{\mu \nu}=2b^2 T_{\mu \nu}~,\label{EoM_G_vec}
\end{align}
where $b=\kappa_g/g_c$ characterizes the strength of the backreaction of the matter fields on the metric. $T_{\mu \nu}$ is the stress-energy tensor of the matter fields with the following expression:
\begin{align}
        T_{\mu \nu}=&\frac{1}{2}g_{\mu \nu}\Big( -\frac{1}{4}F_{\alpha\beta}^a F^{a\alpha\beta}-\frac{1}{4}\tilde{F}_{\alpha\beta} \tilde{F}^{\alpha\beta}-\tilde{D}_{\alpha}\Psi^{a}(\tilde{D}^{\alpha}\Psi^{a})^*-m^{2}|\Psi^{a}|^2 \Big)\nonumber\\
        &+\frac{1}{2}F_{\mu \rho}^a F_{\nu}^{a \rho}+\frac{1}{2}\tilde{F}_{\mu \rho} \tilde{F}_{\nu}^{\,\, \rho}+\frac{1}{2}\Big(\tilde{D}_{\mu} \Psi^{a} (\tilde{D}_{\nu} \Psi^{a})^*+\mu \leftrightarrow\nu\Big)~.\label{Tuv}
\end{align}
Working in the grand canonical ensemble with fixed chemical potential, the grand potential of the system equals the temperature $T$ times the Euclidean on-shell action of the full system, considering the boundary terms as
\begin{align}
	\Omega=TS_{E}=&T \Big[-\frac{1}{2\kappa_g^2}\int d^4x \sqrt{g} \Big[ R+\frac{6}{L^2}+2b^2 \Big(-\frac{1}{4}F^a_{\mu \nu}F^{a\mu \nu}-\frac{1}{4}\tilde{F}_{\mu\nu} \tilde{F}^{\mu\nu}\nonumber \\
    &-\tilde{D}_{\mu}\Psi^a(\tilde{D}^{\mu}\Psi^a)^*-m^2|\Psi^a|^2\Big) \Big]-\frac{1}{\kappa_g^2}\int_{r \to \infty}d^3x\sqrt{h}\left(K-\frac{2}{L} \right) \Big]~.\label{Omega_u1su3}
\end{align}

While Eq.~\eqref{Omega_u1su3} defines the general thermodynamic potential for the system, identifying the favored condensate configuration is technically simpler in the probe limit at zero baryon chemical potential. In the following subsection, we turn to the pure $SU(3)_c$ sector and derive the corresponding probe free energy to systematically screen the condensation channels.

\subsection{Channel selection in the probe limit at zero baryon chemical potential}\label{sec2.2}
Setting the $U(1)$ sector in Eq.~\eqref{S_SU3U1} to zero yields the holographic model with a scalar octet coupled to only the $SU(3)_c$ gauge fields. To compare the stability of different condensation  channels in this model, we perform a comparative analysis in the probe limit, where the background geometry simplifies to a (3+1)-dimensional black brane
\begin{align}
	ds^{2}=-f(r)dt^{2}+\frac{1}{f(r)}dr^{2}+r^2 (dx^{2}+dy^{2})~,\label{metric_probelimit}
\end{align}
with $f(r)=(r^2-r_h^3/r)/L^2$. For simplicity, we set $L=1$. The Hawking temperature of this solution is $T=3r_h/(4\pi)$.

The matter fields yield the equations of motion as follows:
\begin{align}
        g^{\mu \nu} D_{\mu} D_{\nu} \Psi^{a} - m^2 \Psi^{a} = 0~,\label{EQM_psi_SU3_vec}\\
	      g^{\rho \nu} D_{\rho}F_{\nu\mu}^{a}-f^{abc}(\Psi^{b})^*D_{\mu}\Psi^{c}-f^{abc}\Psi^{b}(D_{\mu}\Psi^{c})^*=0~.\label{EQM_A_SU3_vec}
\end{align}

The index $a$ runs from 1 to 8, giving eight equations for both $\Psi^{a}$ and $A_{\mu}^{a}$. We focus on static homogeneous solutions that have no explicit dependence on \{$t$, $x$, $y$\}, and set $A_r^a = 0$. Given the symmetry between $A_x^a$ and $A_y^a$, it is sufficient to activate one of them without loss of generality.

Considering the competition and coexistence between the s-wave and p-wave orders in a similar $SU(2)$ setup~\cite{Nie:2013sda,Nie:2015zia}, it is consistent to work within the single-scalar and single-vector truncation, where both one scalar and one spatial vector are activated simultaneously:
\begin{align}\label{Ansatz}
A^{1}_t=\phi(r)~,\quad\Psi^{a}=\psi_{s}(r)~,\quad  A^{b}_x=\psi_{p}(r)~.
\end{align}
The color indices $a$ and $b$ are preserved for a scan over the color indices within this truncation. Substituting this ansatz into Eqs.~\eqref{EQM_psi_SU3_vec} and \eqref{EQM_A_SU3_vec} yields:
\begin{align}
	\phi''+\frac{2}{r}\phi'-\frac{2q_s^2r^2\psi_s^2+q_p^2\psi_p^2}{r^2 f}\phi=&0~,\label{sp_phi_su3}\\
    \psi_p''+\frac{f'}{f}\psi_p'+\frac{q_p^2\phi^2}{f^2}\psi_p-\frac{2\eta^2\psi_s^2\psi_p}{f}=&0~,\label{sp_psip_su3}\\
	\psi_s''+(\frac{2}{r}+\frac{f'}{f})\psi_s'-\frac{m^2}{f}\psi_s+\frac{q_s^2\phi^2}{f^2}\psi_s-\frac{\eta^2\psi_p^2\psi_s}{r^2f}=&0~,\label{sp_psis_su3}
\end{align}
with the values of the effective couplings \{$q_s$, $q_p$, $\eta$\} listed in Table~\ref{IndexTable}. We can see that although the specific values of the indices $\{a,b\}$ in the ansatz are different, the structure of the equations remains unchanged, and the differences only appear in the values of the effective couplings $q_s$, $q_p$, and $\eta$.

\begin{table}[h]
\centering
\small
\renewcommand{\arraystretch}{1.2}
\setlength{\tabcolsep}{4.5pt}

\caption{Classification of effective coupling parameters across color channels $(a, b)$. The entries list $2(q_s, q_p, \eta)$. The symbol \text{`--'} indicates an uncoupled interaction.}
\label{IndexTable}
\begin{tabular}{ccccccccc}
\toprule
$2(q_s, q_p, \eta)$ & $a=1$ & $2$ & $3$ & $4$ & $5$ & $6$ & $7$ & $8$ \\
\midrule
$b=1$ & $(0,0,0)$ & $(2,0,2)$ & $(2,0,2)$ & $(1,0,1)$ & $(1,0,1)$ & $(1,0,1)$ & $(1,0,1)$ & $(0,0,0)$ \\
$2$   & $(0,2,2)$ & $(2,2,0)$ & $(2,2,2)$ & $(1,2,1)$ & $(1,2,1)$ & $(1,2,1)$ & $(1,2,1)$ & $(0,2,0)$ \\
$3$   & $(0,2,2)$ & $(2,2,2)$ & $(2,2,0)$ & $(1,2,\text{--})$ & $(1,2,\text{--})$ & $(1,2,\text{--})$ & $(1,2,\text{--})$ & $(0,2,0)$ \\
$4$   & $(0,1,1)$ & $(2,1,1)$ & $(2,1,\text{--})$ & $(1,1,0)$ & $(1,1,2)$ & $(1,1,1)$ & $(1,1,1)$ & $(0,1,\text{--})$ \\
$5$   & $(0,1,1)$ & $(2,1,1)$ & $(2,1,\text{--})$ & $(1,1,2)$ & $(1,1,0)$ & $(1,1,1)$ & $(1,1,1)$ & $(0,1,\text{--})$ \\
$6$   & $(0,1,1)$ & $(2,1,1)$ & $(2,1,\text{--})$ & $(1,1,1)$ & $(1,1,1)$ & $(1,1,0)$ & $(1,1,2)$ & $(0,1,\text{--})$ \\
$7$   & $(0,1,1)$ & $(2,1,1)$ & $(2,1,\text{--})$ & $(1,1,1)$ & $(1,1,1)$ & $(1,1,2)$ & $(1,1,0)$ & $(0,1,\text{--})$ \\
$8$   & $(0,0,0)$ & $(2,0,0)$ & $(2,0,0)$ & $(1,0,\text{--})$ & $(1,0,\text{--})$ & $(1,0,\text{--})$ & $(1,0,\text{--})$ & $(0,0,0)$ \\
\bottomrule
\end{tabular}
\end{table}

Solving Eqs.~\eqref{sp_phi_su3}--\eqref{sp_psis_su3} requires appropriate boundary conditions. The expansions of the fields near the horizon are:
\begin{align}
	\phi(r)&=\phi^{h1}(r-r_{h})+\phi^{h2}(r-r_{h})^2+\mathcal{O}((r-r_{h})^{3})~,\nonumber\\
	\psi_s(r)&=\psi_{s}^{h0}+\psi_{s}^{h1}(r-r_{h})+\mathcal{O}((r-r_{h})^2)~,\\
    \psi_p(r)&=\psi_{p}^{h0}+\psi_{p}^{h1}(r-r_{h})+\mathcal{O}((r-r_{h})^2)~.\nonumber
\end{align}
Near the AdS boundary, the expansions take the form:
\begin{align}
	\phi(r)=\mu-\frac{\rho}{r}+\dots~,\quad \psi_s(r)=\frac{\psi_{s-}}{r^{\Delta_{s-}}}+\frac{\psi_{s+}}{r^{\Delta_{s+}}}+\dots~,\quad\psi_p(r)=\psi_{p-}+\frac{\psi_{p+}}{r}+\dots~.
\end{align}

According to the AdS/CFT dictionary, $\mu$ and $\rho$ correspond to chemical potential and charge density, respectively. $\{\psi_{s-}, \psi_{p-}\}$ and $\{\psi_{s+}, \psi_{p+}\}$ correspond to the sources and expectation values. In this paper, standard quantization is adopted, and the source is set to zero to study spontaneous symmetry breaking in the model. $\Delta_{s-}$ and $\Delta_{s+}$ are the conformal dimensions controlled by the mass parameter $m^2$ as $\Delta_{s\pm}=(3 \pm \sqrt{9+4m^2})/2$.

We work in the grand canonical ensemble. The grand potential is identified with the Euclidean on-shell action of the bulk solution multiplied by the temperature. In the probe limit, we need only to compare the contribution from the matter fields:
\begin{align}
	\Omega_m=&TS_{\text{ME}}=\frac{V_2}{g_c^2} (-\frac{\mu \rho}{2}+\int_{r_h}^{\infty} (\frac{q_s^2 r^2 \phi^2 \psi_s^2}{f}+\frac{q_p^2 \phi^2 \psi_p^2}{2f}+\eta^2\psi_s^2\psi_p^2)dr)~,\label{Omega_probelimit_su3}
\end{align}
where $S_{\text{ME}}$ denotes the Euclidean on-shell action of the matter sector on the black brane background, and $V_2$ is the area of the two-dimensional transverse space.

\begin{figure}[h]
\centering
\includegraphics[width=0.48\linewidth]{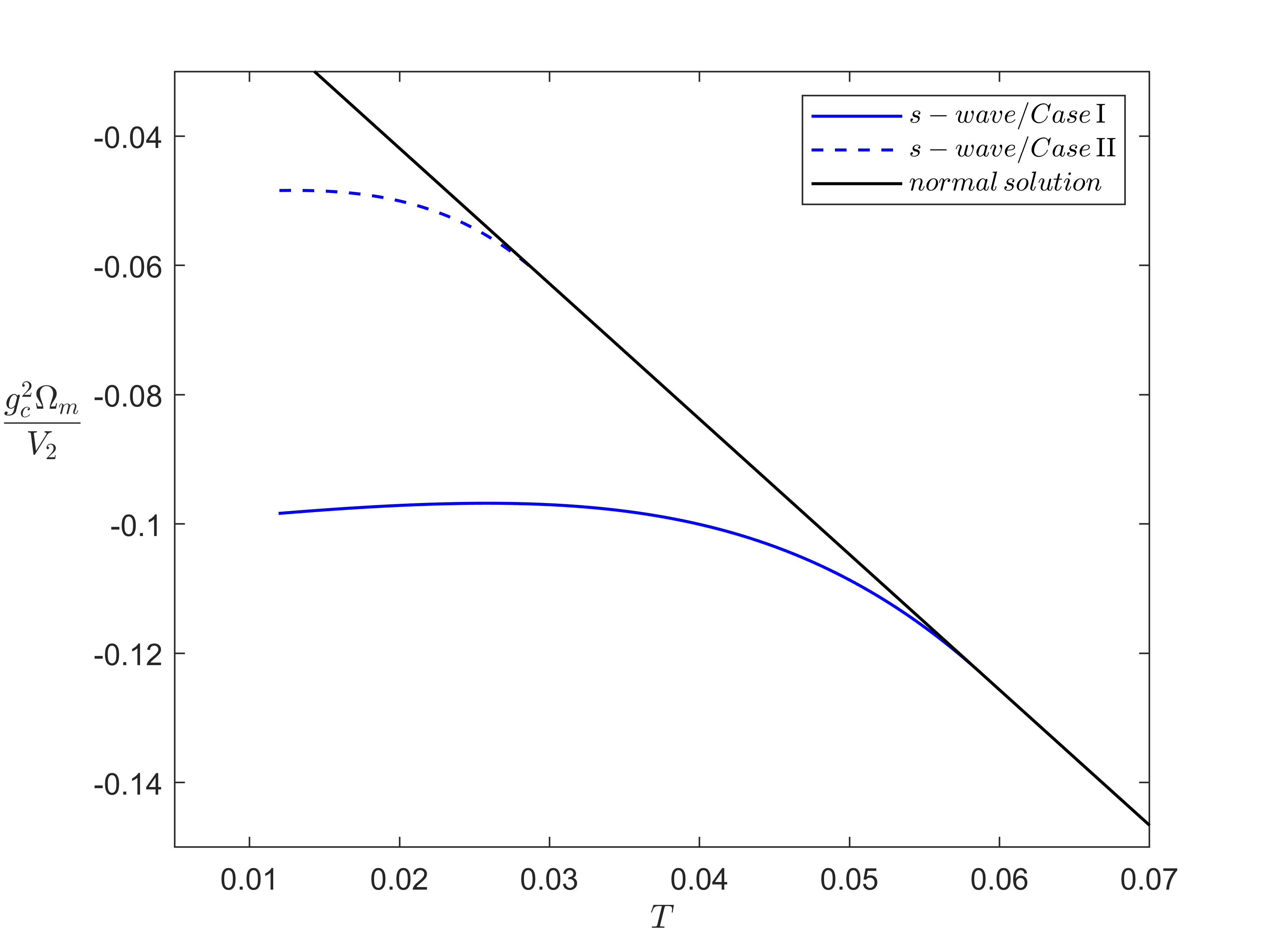}
\includegraphics[width=0.48\linewidth]{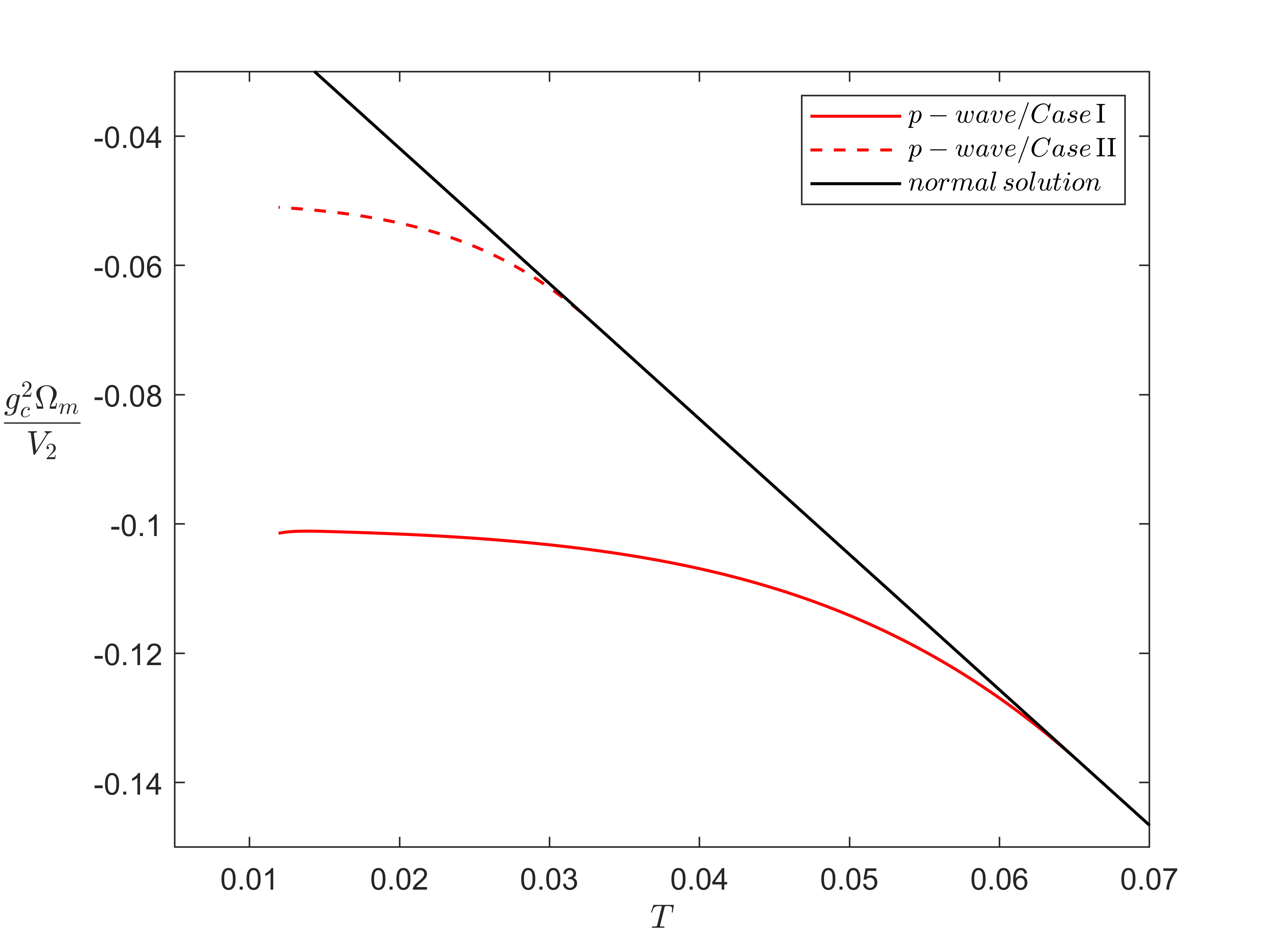}
\caption{The grand potential curves of the single condensate s-wave \textbf{(Left)} and p-wave \textbf{(Right)} solutions with different values of $q_s$ and $q_p$. The detailed values of the charge couplings are $q_s=q_p=1$ for Case I, $q_s=q_p=1/2$ for Case II.}\label{fig1}
\end{figure}

From Table~\ref{IndexTable}, we can see that there are only three possible values \{0, 1/2, 1\} for each of the effective couplings \{$q_s$, $q_p$, $\eta$\}. $q_s$ (or $q_p$) describes the effective charge coupling between the s-wave (or the p-wave) order and the gauge potential $A_t^1$, and $\eta$ describes the coupling between the s-wave and p-wave order. When either $a$ or $b$ equals 1 or 8, the relative order $\Psi^a$ or $A^b_x$ decouples from $A_t^1$ and will not condense. For example, the case with $a=b=1$ (or $a=b=8$) will not show any condensate, and the normal phase is always the most stable. The case with a single condensate is a bit more complicated and interesting. There are two possible nonzero values, $1/2$ and $1$, for the effective charge of the single order. According to the well-studied results, the larger charge coupling yields more stable condensate solutions, which also exhibit lower values of the grand potential density $\Omega_m$, as illustrated in Figure~\ref{fig1}. Therefore, the choices with $q_s=1$ or $q_p=1$ win the competition within the scanned channels for the single-condensate s-wave or p-wave solutions. We can also see from the table that, given $ q_s=1$ or $ q_p=1$, the choice of $a$ or $b$ is not unique; therefore, there are degenerate structures for the s-wave and p-wave solutions. In order to get the most stable single-condensate s-wave and p-wave solutions, we should constrain the values of $a$ and $b$ to those making $q_s=q_p=1$.

The value of $\eta$ controls the coupling between the s-wave and p-wave order, which is important in determining the stability of the coexistent solutions but does not change the single condensate solutions. From the knowledge of the s+p models~\cite{Nie:2013sda,Zhang:2021vwp}, the lower value of $\eta$ makes the s+p solution more stable and the s+p solutions with $\eta=0$ are more stable than the single condensate s-wave and p-wave phases. Therefore, based on the above constraint of $q_s=q_p=1$ for the single condensate solutions, we further choose the combination of $a$ and $b$ such that $\eta=0$ to get the most stable coexistent solutions. Finally, only two choices $a=b=2$ and $a=b=3$ are left for further investigation. These two choices yield equally stable solutions and admit more complex degenerate solutions. We take $a=b=2$ to study the phase structure at $\mu_b=0$, which does not cover the complex degenerate solutions, but the s-wave, p-wave, and s+p labels in the phase diagram are the same as those in a more general setup.

With $a=b=2$, we get $q_s=q_p=1$ and $\eta=0$ in the equations of motion, which become the same as the equations of motion in the $SU(2)$ holographic model~\cite{Nie:2013sda}. Therefore, we also get exactly the same $\Delta-T$ phase diagram presented in Ref.~\cite{Nie:2013sda}.

Consequently, within this truncation in the absence of the $U(1)$ sector, the model reduces to the well-studied $SU(2)$ s+p model ~\cite{Nie:2013sda, Nie:2014qma,Nie:2015zia}. This provides a useful check of the retained channel once the baryon chemical potential is turned on.

\section{Color Superconductivity at Finite Baryon Chemical Potential}\label{sec3}
We now proceed to investigate the more complicated condensates at finite baryon chemical potential.

\subsection{Equations of motion for the retained ansatz at finite baryon chemical potential with backreaction}\label{sec3.1}
When returning to the full model, the metric ansatz for the hairy black brane is
\begin{align}
ds^2=&-N(r)\sigma(r)^2dt^2+\frac{dr^2}{N(r)}+\frac{r^2}{f(r)^2}dx^2+r^2f(r)^2dy^2~,\label{metric}
\end{align}
with $N(r)=(r^2-2M(r)/r)/L^2$, and the Hawking temperature is $T=N'(r_h) \sigma(r_h)/ (4\pi)$.

To study the phase diagram at finite baryon chemical potential $\mu_b$, we activate the $U(1)$ gauge field which couples $\Psi^{2}$ with $\Psi^{3}$. Here, consistency of the equations of motion requires $\Psi^{3}$ to be activated as in the following ansatz:
\begin{align}
	&A^{1}_t=\phi_3(r)~,\quad \tilde{A}_t=\phi_{1}(r)~,\quad A^{2}_x=\psi_{p}(r)~, \quad \Psi^{2}=\psi_{s1}(r)~,\quad \Psi^{3}=i\psi_{s2}(r)~.\label{ansatzCSC}
\end{align}

The scalar sector is generalized to an $S_1+iS_2$ structure. The equations of motion are presented as follows.
\begin{align}
        &\phi_1''+(\frac{2}{r}-\frac{\sigma'}{\sigma})\phi_1'-\frac{8\phi_1 }{9 N}(\psi_{s1}^2+\psi_{s2}^2)-\frac{8\phi_3 \psi_{s1} \psi_{s2}}{3 N}=0~,\label{EoM_At_U1}\\
        &\phi_3''+(\frac{2}{r}-\frac{\sigma'}{\sigma})\phi_3'-\frac{2\phi_3 }{N}(\psi_{s1}^2+\psi_{s2}^2)-\frac{8\phi_1 \psi_{s1} \psi_{s2}}{3 N}-\frac{f^2\psi_{p}^2\phi_3}{r^2N}=0~,\\
        &\psi_{p}''+(\frac{\sigma'}{\sigma}+\frac{N'}{N}+\frac{2f'}{f})\psi_{p}'+\frac{\phi_3^2 \psi_{p}}{N^2 \sigma^2}-\frac{2\psi_{s2}^2\psi_{p}}{N}=0~,\\
        &\psi_{s1}''+(\frac{2}{r}+\frac{\sigma'}{\sigma}+\frac{N'}{N})\psi_{s1}'-\frac{m^2\psi_{s1}}{N}+\frac{4\phi_1^2 \psi_{s1}}{9N^2 \sigma^2}+\frac{\phi_3^2 \psi_{s1}}{N^2 \sigma^2}+\frac{4\phi_3 \phi_1 \psi_{s2}}{3N^2 \sigma^2}=0~,\label{EoM_psis1_U1}\\
        &\psi_{s2}''+(\frac{2}{r}+\frac{\sigma'}{\sigma}+\frac{N'}{N})\psi_{s2}'-\frac{m^2\psi_{s2}}{N}+\frac{4\phi_1^2 \psi_{s2}}{9N^2 \sigma^2}+\frac{\phi_3^2 \psi_{s2}}{N^2 \sigma^2}+\frac{4\phi_3 \phi_1 \psi_{s1}}{3N^2 \sigma^2}-\frac{f^2\psi_{p}^2\psi_{s2}}{r^2N}=0~,\label{EoM_psis2_U1}\\
        &f''+\frac{f'\sigma'}{\sigma}+\frac{f'N'}{N}-\frac{f'^2}{f}+\frac{2f'}{r}-\frac{b^2f^3\psi_{p}'^2}{2r^2}+\frac{b^2f^3\phi_3^2\psi_{p}^2}{2r^2N^2\sigma^2}-\frac{b^2f^3\psi_{p}^2\psi_{s2}^2}{r^2N}=0~,\\
        &\sigma'-\frac{r\sigma f'^2}{f^2}-b^2r\sigma(\psi_{s1}'^2+\psi_{s2}'^2)-\frac{4b^2r \phi_1^2 (\psi_{s1}^2+\psi_{s2}^2)}{9N^2 \sigma}-\frac{b^2r\phi_3^2(\psi_{s1}^2+\psi_{s2}^2)}{N^2\sigma}\nonumber\\
        &-\frac{8b^2r\phi_1 \phi_3\psi_{s1}\psi_{s2}}{3N^2\sigma} -\frac{b^2f^2\sigma\psi_{p}'^2}{2r}-\frac{b^2f^2\phi_3^2\psi_{p}^2}{2rN^2\sigma}=0~,\\
        &M'-\frac{r^2Nf'^2}{2f^2}-\frac{b^2r^2(\phi_1'^2+\phi_3'^2)}{4\sigma^2}-\frac{1}{2}b^2m^2r^2(\psi_{s1}^2+\psi_{s2}^2)-\frac{1}{2}b^2r^2N(\psi_{s1}'^2+\psi_{s2}'^2) \nonumber \\
        &-\frac{2b^2r^2\phi_1^2(\psi_{s1}^2+\psi_{s2}^2)}{9N\sigma^2}-\frac{b^2r^2\phi_3^2(\psi_{s1}^2+\psi_{s2}^2)}{2N\sigma^2}-\frac{4b^2r^2\phi_1\phi_3\psi_{s1}\psi_{s2}}{3N\sigma^2}-\frac{1}{4}b^2f^2N\psi_{p}'^2\nonumber\\
        &-\frac{b^2f^2\phi_3^2\psi_{p}^2}{4N\sigma^2}-\frac{1}{2}b^2f^2\psi_{p}^2\psi_{s2}^2=0~.\label{EoM_M}
\end{align}

\subsection{Decoupling transformation, effective chemical potentials, and scaling symmetries}\label{sec3.2}
The finite $\phi_3$ and $\phi_1$ couple $\psi_{s1}$ and $\psi_{s2}$ at the linear level, therefore, the two scalar components are coupled and generically condense together. We find that the recombination $\psi_{s+}=\psi_{s1}+\psi_{s2}$ and $\psi_{s-}=\psi_{s1}-\psi_{s2}$ unties the direct coupling between $\psi_{s1}$ and $\psi_{s2}$ in the absence of the p-wave order, and Eqs.~(\ref{EoM_psis1_U1}) and (\ref{EoM_psis2_U1}) are transformed into
\begin{align}
        &\psi_{s+}''+(\frac{2}{r}+\frac{\sigma'}{\sigma}+\frac{N'}{N})\psi_{s+}'-\frac{m^2}{N}\psi_{s+} +\frac{(\phi_3+2\phi_1/3)^2}{N^2 \sigma^2}\psi_{s+} -\frac{f^2\psi_{p}^2(\psi_{s+}-\psi_{s-})}{2r^2N}=0~,\label{EoMSplus}\\
        &\psi_{s-}''+(\frac{2}{r}+\frac{\sigma'}{\sigma}+\frac{N'}{N})\psi_{s-}'-\frac{m^2}{N}\psi_{s-} +\frac{(\phi_3-2\phi_1/3)^2}{N^2 \sigma^2}\psi_{s-} +\frac{f^2\psi_{p}^2(\psi_{s+}-\psi_{s-})}{2r^2N}=0~.\label{EoM_M1}
\end{align}
With these two new equations, we are able to study the single condensate $S_+$ solutions with $\psi_{s1}=\psi_{s2}$ and the coexistent $S_+ + S_-$ solutions, which means $\psi_{s1}\neq\psi_{s2}$.

Remarkably, the gauge fields enter these decoupled equations exclusively through the effective gauge combinations
\begin{equation}
    \phi_{\text{eff},\pm}(r) \equiv \phi_3(r) \pm \frac{2}{3} \phi_1(r)~.
    \label{EOM_eff_gauge}
\end{equation}
Notice that this exact algebraic property will make it possible to get analytical expressions for the critical temperatures of the $S_+$ and $S_-$ condensate.

Physically, the emergence of $\phi_{\text{eff},\pm}(r)$ demonstrates that the non-Abelian color density and the Abelian baryon density do not act independently. In particular, the enhanced effective potential $\phi_{\text{eff},+}$ significantly promote the condensate of the $\psi_{s+}$ mode, thereby playing the dominant driving role in triggering color superconductivity at finite density.

We identify four sets of scaling symmetries in the full equations which are useful in numerical work.
\begin{align}
        (1)&\quad L \to \lambda^{-1} L~, \, b \to \lambda^{-1} b~, \, \sigma \to \lambda^{-1} \sigma~, \, m \to \lambda m~,\, N \to \lambda^2 N~, \nonumber\\
        &\quad \phi_1 \to \lambda \phi_1~,\,\phi_3 \to \lambda \phi_3~, \, \psi_{s1} \to \lambda \psi_{s1}~, \, \psi_{s2} \to \lambda \psi_{s2}~, \, \psi_p \to \lambda \psi_p~;\label{scaling1}\\
        (2)&\quad r \to \lambda r~, \, M \to \lambda^3 M~,\, N \to \lambda^2 N~, \,  \phi_1 \to \lambda \phi_1~,\,\phi_3 \to \lambda \phi_3~, \, \psi_p \to \lambda \psi_p~;\label{scaling2}\\
        (3)&\quad \phi_1 \to \lambda \phi_1~, \, \phi_3 \to \lambda \phi_3~, \, \sigma \to \lambda \sigma~;\label{scaling3}\\
        (4)&\quad \psi_p \to \lambda \psi_p \, f \to \lambda^{-1} f~.\label{scaling4}
\end{align}

\subsection{Condensation behaviors}\label{sec3.3}
To solve the coupled system of ordinary differential equations numerically, we implement the shooting method with a high-precision adaptive Runge-Kutta-Fehlberg scheme. Near the horizon, the functions admit the expansions:
\begin{alignat}{2}
        \phi_1(r)=&\phi_{1}^{h1}(r-r_h)+\phi_{1}^{h2}(r-r_h)^2+\dots~,&\quad \phi_3(r)&=\phi_{3}^{h1}(r-r_h)+\phi_{3}^{h2}(r-r_h)^2+\dots~,\nonumber \\
        \psi_p(r)=&\psi_{p}^{h0}+\psi_{p}^{h1}(r-r_h)+\dots~,&\quad \psi_{s1}(r)&=\psi_{s1}^{h0}+\psi_{s1}^{h1}(r-r_h)+\dots~,  \\
        \psi_{s2}(r)=&\psi_{s2}^{h0}+\psi_{s2}^{h1}(r-r_h)+\dots~,&\quad  \sigma(r)&=\sigma^{h0}+\sigma^{h1}(r-r_h)+\dots~,\nonumber\\
        f(r)=&f^{h0}+f^{h1}(r-r_h)+\dots~,& \quad M(r)&=r_h^3/2+M^{h1}(r-r_h)+\dots~.\nonumber
\end{alignat}
Only the leading coefficients $\{\phi_{1}^{h1}, \phi_{3}^{h1}, \psi_{p}^{h0}, \psi_{s1}^{h0}, \psi_{s2}^{h0}, \sigma^{h0}, f^{h0}\}$ are independent.

Near the AdS boundary,
\begin{alignat}{3}
        \phi_1(r)=&\mu_{b}-\frac{\rho_{b}}{r}+\dots~, &\quad \phi_3(r)&=\mu_{3}-\frac{\rho_{3}}{r}+\dots~, &\quad \psi_p(r)=&\psi_{p-}+\frac{\psi_{p+}}{r}+\dots~, \nonumber \\
        \psi_{s1}(r)=&\frac{\psi_{s1-}}{r^{\Delta_{-}}}+\frac{\psi_{s1+}}{r^{\Delta_{+}}}+\dots~, &\quad \psi_{s2}(r)&=\frac{\psi_{s2-}}{r^{\Delta_{-}}}+\frac{\psi_{s2+}}{r^{\Delta_{+}}}+\dots~, &\quad  \sigma(r)=&\sigma_{b0}+\frac{\sigma_{b3}}{r^3}+\dots~,\\
         f(r)=&f_{b0}+\frac{f_{b3}}{r^3}+\dots~,  &\quad M(r)&=M_{b0}+\frac{M_{b1}}{r}+\dots~,\nonumber
\end{alignat}
where $\Delta_{\pm}=(3 \pm \sqrt{9+4m^2})/2$. 

We set $\sigma_{b0}=1$ and $f_{b0}=1$ so that the boundary geometry is asymptotically AdS. According to the AdS/CFT dictionary in the standard quantization scheme, $\{\mu_b, \mu_3\}$ and $\{\rho_b, \rho_3\}$ are dual to the chemical potentials and charge densities, respectively, while $\{\psi_{p-}, \psi_{s1-}, \psi_{s2-}\}$ and $\{\psi_{p+}, \psi_{s1+}, \psi_{s2+}\}$ correspond to the sources and expectation values of the dual operators. The recombination of $\psi_{s1}$ and $\psi_{s2}$ also yields the recombination of the dual scalar operators as $\mathcal{O}_{s+}=\mathcal{O}_{s1}+\mathcal{O}_{s2}$ and $\mathcal{O}_{s-}=\mathcal{O}_{s1}-\mathcal{O}_{s2}$.

For the normal phase,
\begin{align}
        \phi_1(r)=&\mu_b(1-\frac{r_h}{r})~, \quad \phi_3(r)=\mu_3(1-\frac{r_h}{r})~,\quad \sigma(r)=1~, \quad f(r)=1~,\nonumber\\
        N(r)=&r^2(1-\frac{r_h^3}{r^3})+b^2\frac{\mu_3^2r_h^2}{2r^2}(1-\frac{r}{r_h})+b^2\frac{\mu_b^2r_h^2}{2r^2}(1-\frac{r}{r_h})~.\label{normal_backreaction}
\end{align}
The temperature and free energy are
\begin{align}
        T=\frac{r_h}{4\pi}(3-\frac{b^2\mu_b^2}{2r_h^2}-\frac{b^2\mu_3^2}{2r_h^2})~, \quad \frac{2\kappa_g^2}{V_2}\Omega=-r_h^3-\frac{1}{2}b^2\mu_b^2r_h-\frac{1}{2}b^2\mu_3^2r_h~. \label{T_Omega_normal_back}
\end{align}
For the condensed phases, we have
\begin{align}
        &T=\frac{r_h}{4\pi}(3\sigma^{h0}-\frac{b^2{\phi_1^{h1}}^2}{2\sigma^{h0}}-\frac{b^2{\phi_3^{h1}}^2}{2\sigma^{h0}}-b^2m^2\sigma^{h0}({\psi_{s1}^{h0}}^2+{\psi_{s2}^{h0}}^2)-b^2 {f^{h0}}^2\sigma^{h0}{\psi_{s2}^{h0}}^2{\psi_{p}^{h0}}^2)~, \nonumber\\ 
        &\frac{2\kappa_g^2}{V_2}\Omega=\lim_{r \to \infty} \Big[ \frac{2r^2N\sigma f'}{f}-r^2\sigma N'-2r^2 N \sigma'+4r^2\sqrt{N}\sigma-2rN\sigma \Big]=-2M_{b0}~. \label{T_Omega_conden_back}
\end{align}

In the probe limit with $b=0$, we need only calculate the contribution of the matter fields to the grand potential for the various solutions:
\begin{align}
	\Omega_m=TS_{\text{ME}}=\frac{V_2}{g_c^2} \Big(&-\frac{\mu_3 \rho_3}{2}-\frac{\mu_b \rho_b}{2}+\int_{r_h}^{\infty} \big(\frac{r^2 \phi_3^2 (\psi_{s1}^2+\psi_{s2}^2)}{N(r)\sigma(r)}+\frac{4r^2 \phi_1^2 (\psi_{s1}^2+\psi_{s2}^2)}{9N(r)\sigma(r)}\nonumber\\
    &+\frac{8r^2 \phi_1 \phi_3 \psi_{s1}\psi_{s2}}{3N(r)\sigma(r)}+\frac{f^2\phi_3^2 \psi_p^2}{2N(r)\sigma(r)}-f^2\sigma(r)\psi_p^2\psi_{s2}^2\big)dr\Big)~,\label{Omega_probelimit_su3u1}
\end{align}
where $S_{\text{ME}}$ denotes the matter sector of the Euclidean action on the fixed black brane background, and $V_2$ is the area of the two-dimensional transverse space.

For numerical convenience, we set the AdS radius $L=1$ and the horizon radius $r_h=1$. The physical dependence on $r_h$ is later restored using the system's scaling symmetry to obtain a temperature dependence. In order to build the $\mu_b$–$T$ phase disagram, we define the dimensionless ratio $\mu_b/\mu_3$ and set $\mu_3=1$, which reduces the ratio to $\mu_b/\mu_3=\mu_b$ as shown in the figures.

\begin{figure}[h]
\centering
\includegraphics[width=0.48\linewidth]{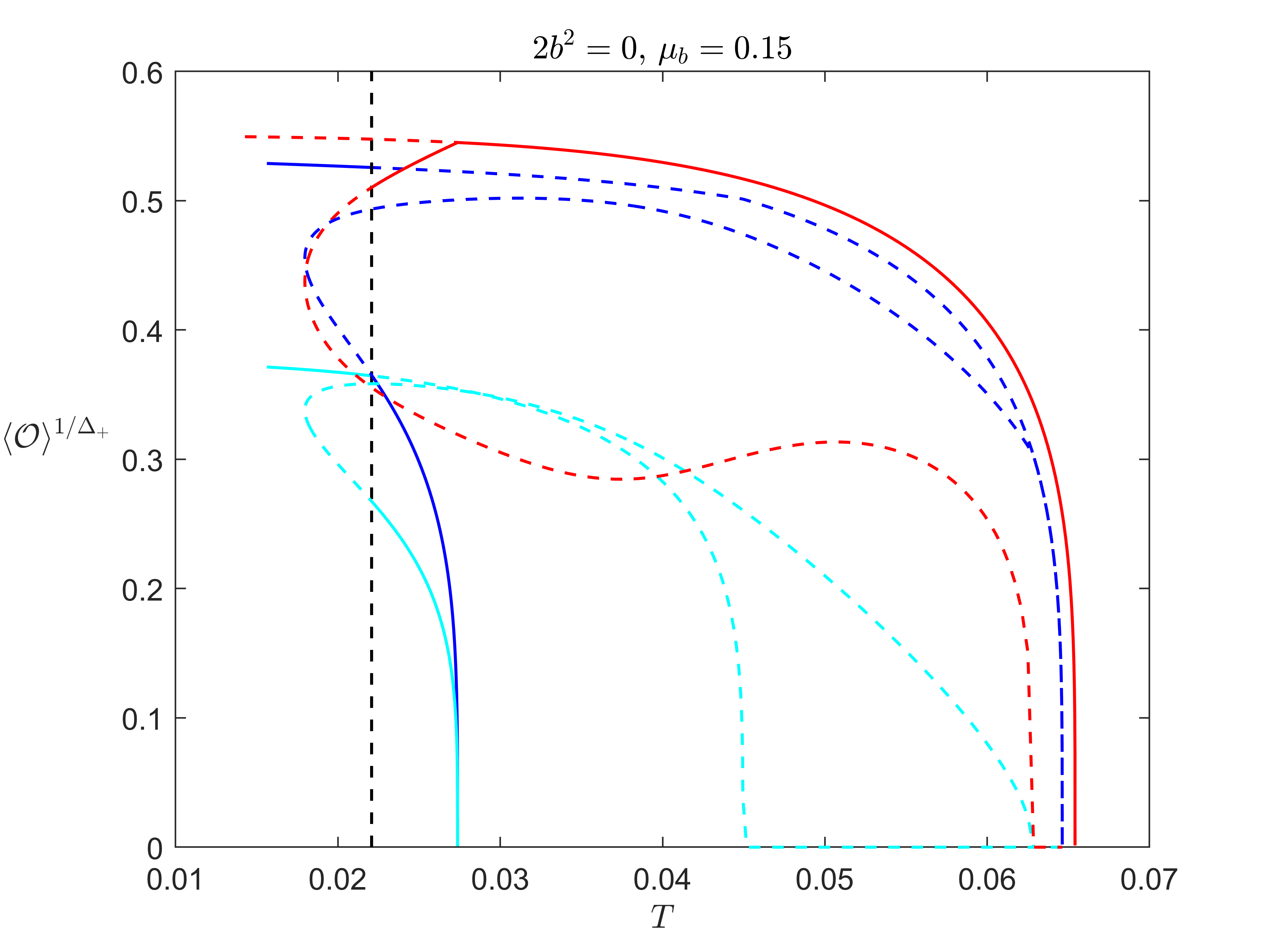}
\includegraphics[width=0.48\linewidth]{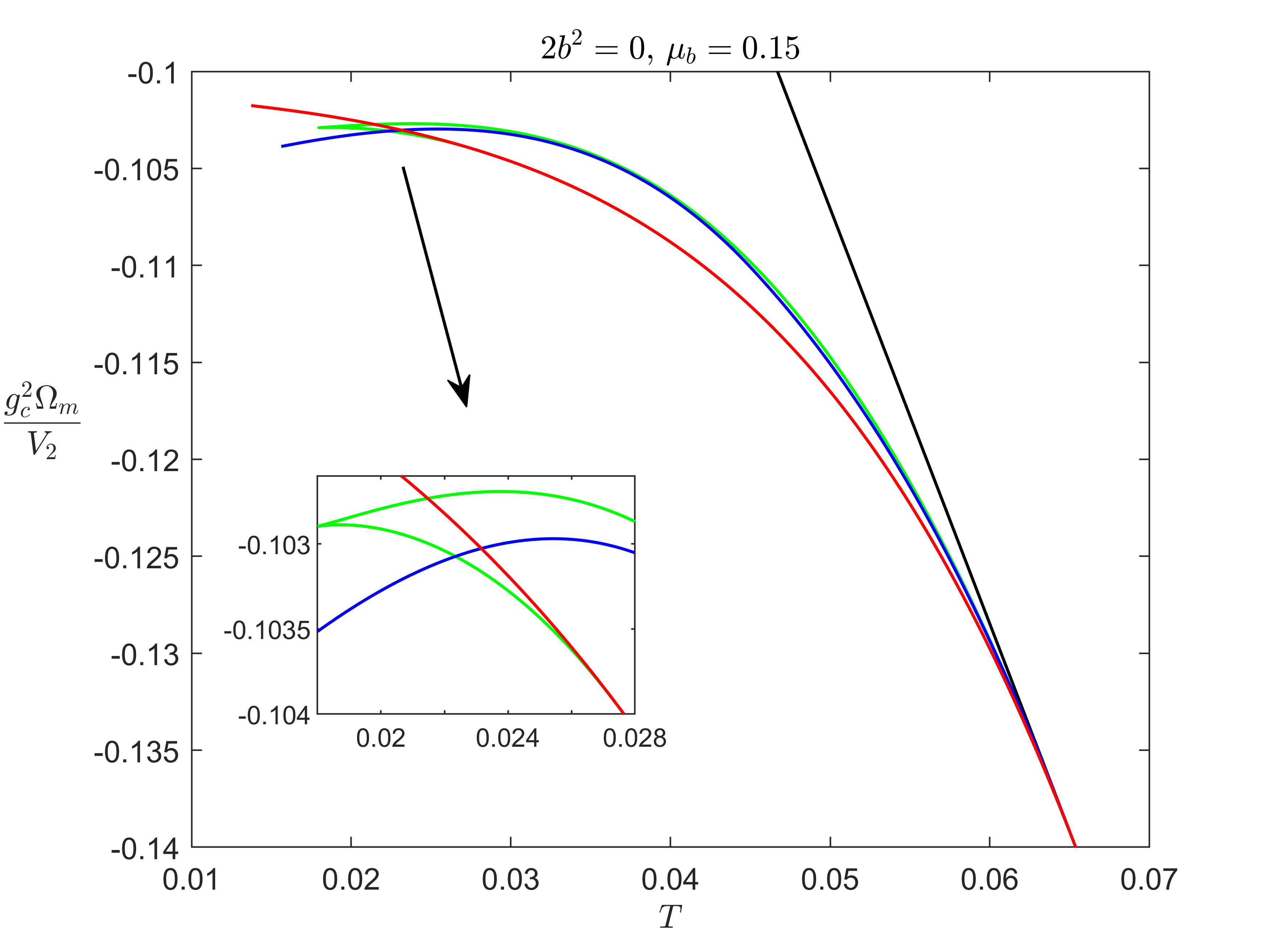}
\caption{\textbf{Left:} Condensate as a function of temperature for the orders $\mathcal{O}_{s+}$, $\mathcal{O}_{s-}$ and $\mathcal{O}_{p}$ with $2b^2=0$ and $\mu_b=0.15$. \textbf{Right:} The grand potential curve for the various solutions at $2b^2=0$ and $\mu_b=0.15$. In the left panel, the blue, cyan, and red curves denote the $\mathcal{O}_{s+}$, $\mathcal{O}_{s-}$, and $\mathcal{O}_{p}$ condensates, respectively. In the right panel, the blue, red, and green curves represent the $S_1+iS_2$, p-wave, and $S_1+iS_2$+p solutions, respectively.}\label{fig2}
\end{figure}

In Figure~\ref{fig2}, we illustrate the condensates and grand potential densities for various solutions in a typical case with $b=0$ (probe limit) and $\mu_b=0.15$, where $b=\kappa_g/g_c$ characterizes the backreaction strength. In the left panel, solid (dashed) curves denote thermodynamically stable (unstable) solutions. The vertical black dashed line marks a first-order phase transition point. We can see that the p-wave phase is favorer at higher temperatures, while the $S_1+iS_2$ phase takes over at very low temperatures. Between them, an intermediate region hosts the coexistent $S_1+iS_2$+p phase. Notably, the $S_1+iS_2$ order generically exhibits unbalanced $S_1$ and $S_2$ components in both the low-temperature and coexistent phases. The balanced $S_1+iS_2$ phase with $\mathcal{O}_{s1}=\mathcal{O}_{s2}$ appears only in the higher-temperature region in the absence of the $S_-$ condensate denoted by the cyan curves, and is a unstable branch in this concrete example.

\section{Phase Diagrams}\label{sec4}
With the equations of motion and the formula for grand potential, we are now in a position to explore the global phase structure and the competitive dynamics between different superconducting orders in the $\mu_b$–$T$ plane. 

\subsection{Phase diagram and analytical scaling in the probe limit}\label{sec4.1}
\begin{figure}[h]
\centering
\includegraphics[width=0.6\linewidth]{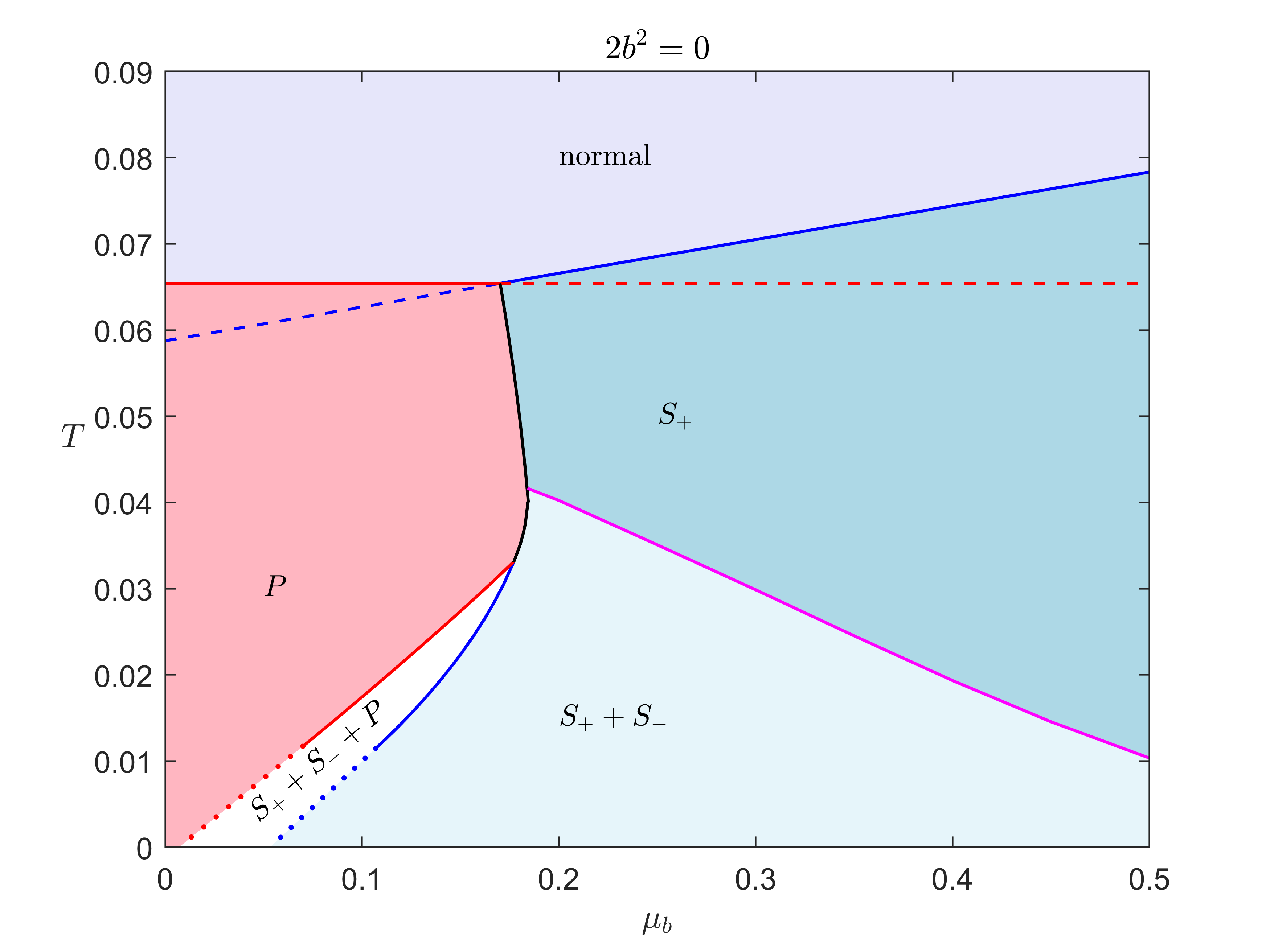}
\caption{$\mu_b$–$T$ phase diagram for $2b^2=0$. The colored regions represent the normal phase (purple), the p-wave phase (red), the $S_+$ phase (blue), the $S_+ + S_-$ phase (light blue), and the $S_+ + S_-$+p phase (white region). The phase boundaries are denoted by solid curves, while the dashed curves indicate the boundaries between unstable solutions. The dotted lines indicate extrapolations where direct computation becomes difficult near $T\rightarrow 0$.}\label{fig3}
\end{figure}

The phase structure of the model is governed by the competition and coexistence between the $S_1+iS_2$ and p-wave orders. Holding $2b^2=0$ and varying $\mu_b$, we construct the $\mu_b$–$T$ phase diagram in the probe limit (Figure~\ref{fig3}). At high temperatures, the system is in the normal phase without condensates. Upon cooling, it undergoes a second-order phase transition to either the p-wave phase (at low $\mu_b$) or the $S_+$ phase (at high $\mu_b$) with balanced $S1$ and $S2$ orders. The $S_+ + S_-$ phase with unbalanced $S_1$ and $S_2$ condensates appears at low temperatures and low $\mu_b$. The p-wave order is favored at low $\mu_b$.

In the probe limit, a key simplification arises at the critical points, where the condensate vanishes, so the metric functions reduce to those of the neutral black brane,
\begin{align}
      N(r)=\frac{r^2}{L^2}(1-(\frac{r_h}{r})^3)~, \quad \sigma(r)=1~, \quad f(r)=1~.\label{metric_fun_probelimit}
\end{align}
In the single-condensate p-wave solution, $\tilde{A}_t$ does not enter the p-wave equation, hence $T_c^p$ is independent of $\mu_b$, resulting in a horizontal phase boundary for the p-wave phase in Figure~\ref{fig3}.

For the $S_1+iS_2$ phase, the equations couple both $\tilde{A}_t$ and $A_t^1(r)$ to the $S_+$ order in a symmetric combination $\phi_{\text{eff}, +}(r)$. This suggests an effective chemical potential
\begin{align}
        \mu^{\text{eff}}_{+}=\mu_{3}+\frac{2}{3}\mu_{b}~,\label{critical_mu}
\end{align}
which determines the critical temperature of the $S_+$ order $T_{c}^{S+}$. Using scaling symmetry, we derive an analytic expression:
\begin{align}
T_{c}^{S_+}=T_c^{0}\left(1+\frac{2}{3}\frac{\mu_b}{\mu_3}\right)~,\label{critical_T}
\end{align}
where $T_c^{0}$ is the critical temperature of the $S_+$ order at $\mu_b=0$. Thus, $T_c^{S_+}$ increases linearly with $\mu_b$ in the probe limit.

\subsection{Effects of backreaction and phase diagrams}\label{sec4.2}
\begin{figure}[h]
\centering
\includegraphics[width=0.48\linewidth]{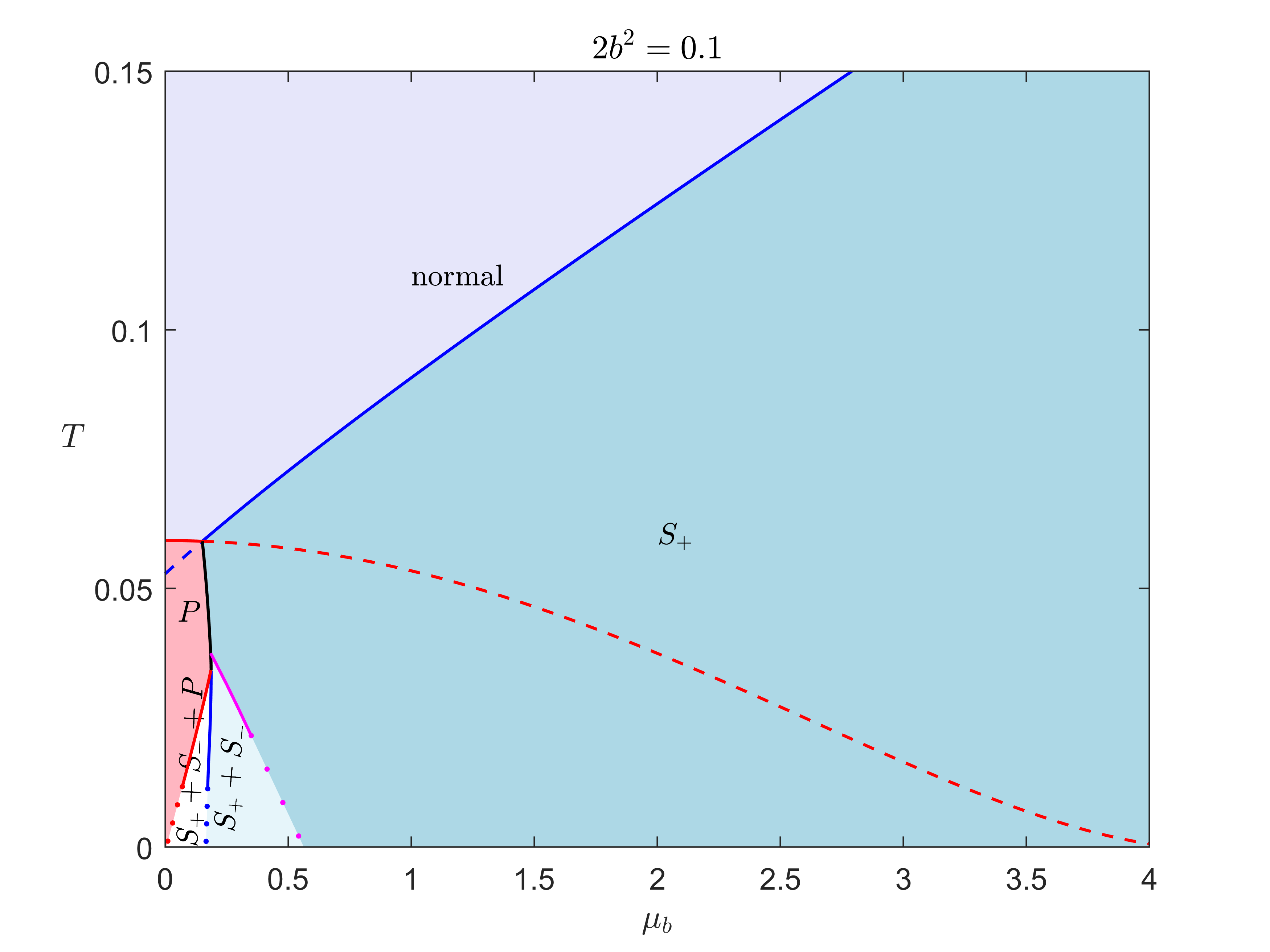}
\includegraphics[width=0.48\linewidth]{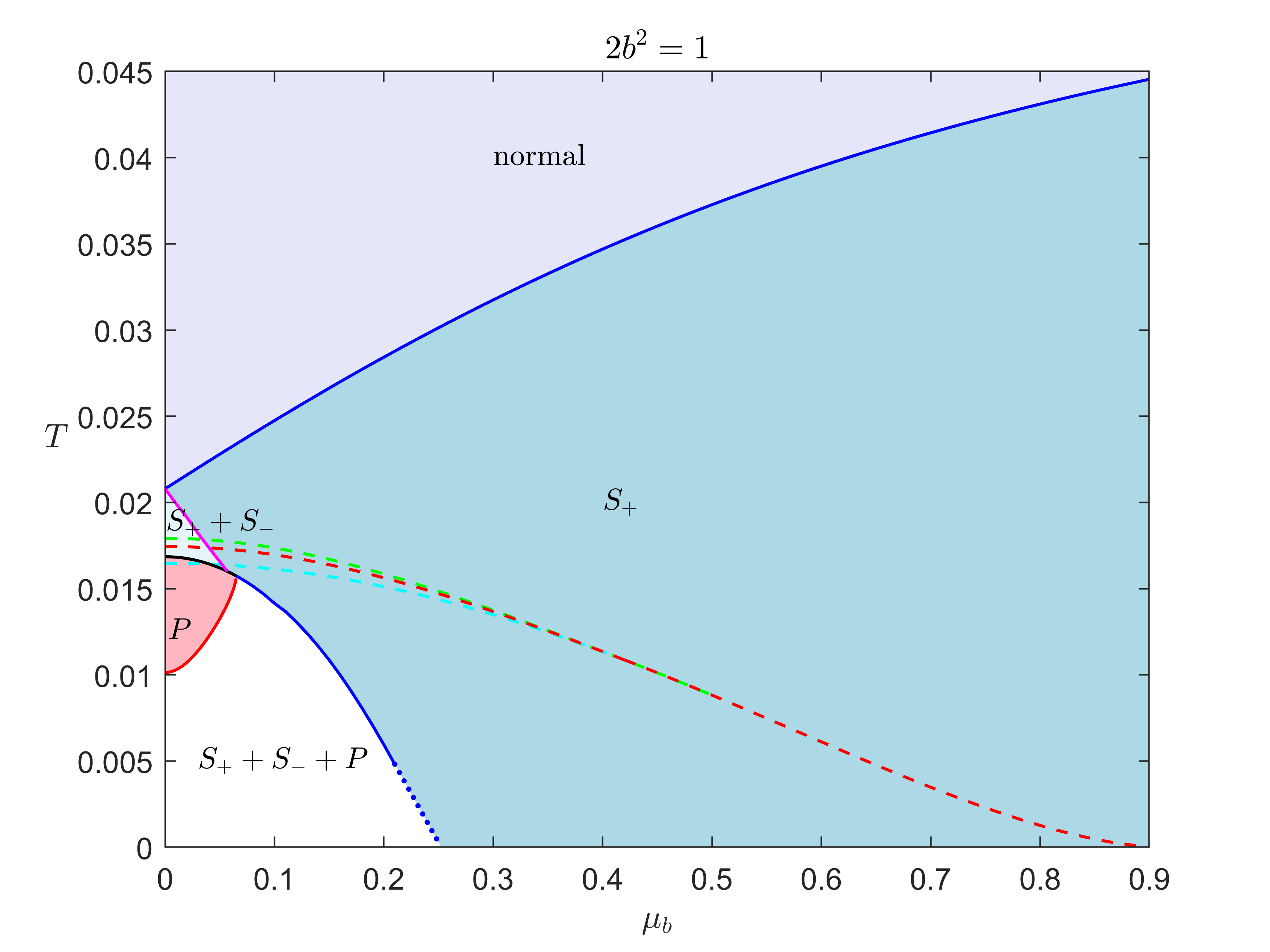}
\caption{$\mu_b$–$T$ phase diagrams for different values of the backreaction parameter $b$ with $2b^2=0.1$ \textbf{(Left)} and $1$ \textbf{(Right)}. Specifically, the red dashed line in the right panel marks the phase transition points of the single-condensate p-wave first-order phase transition, with its quasi-critical points indicated by the cyan and green dashed lines.}\label{fig4}
\end{figure}

When backreaction on the background metric is included ($2b^2>0$), the phase boundaries undergo modifications, as illustrated in Figure~\ref{fig4} for $2b^2=0.1$ (left panel) and $2b^2=1$ (right panel). 
Most notably, the linear scaling relation for the scalar critical temperature, as given in Eq.~\eqref{critical_T}, which holds strictly in the probe limit, is noticeably broken and deformed into a nonlinear bending trajectory. 

To understand why this linear scaling fails despite the scalar matter sector retaining the identical effective gauge combination $\phi_{\text{eff},+} = \phi_3 + 2\phi_1/3$, we inspect the thermodynamic responses of the system. In the probe limit ($b=0$), the background is a neutral Schwarzschild-$\text{AdS}_4$ black brane whose Hawking temperature $T=3r_h/(4\pi)$ is independent of the chemical potentials. However, once backreaction is turned on, the two gauge fields affect the Hawking temperature:
\begin{equation}
    T = \frac{r_h}{4\pi}\left( 3 - \frac{b^2 \mu_3^2}{2r_h^2} - \frac{b^2 \mu_b^2}{2r_h^2} \right)~.\label{HT}
\end{equation}
This temperature suppression breaks the scaling symmetry of the system, thereby giving rise to the nonlinear curvature of the $S_+$ phase boundary.

Furthermore, the competition between the vector and scalar orders is significantly altered by the backreaction. As shown in Figure~\ref{fig4}, the critical temperature of the p-wave condensate $T_c^p$ is no longer independent of $\mu_b$. Instead, it monotonically decreases with increasing baryon density and eventually terminates at a finite $\mu_b$. This is because as $\mu_b$ increases, the temperature \eqref{HT} of the system decreases. While the scalar mode $\psi_{s+}$ benefits directly from the effective potential $\phi_{\text{eff},+}=\phi_3+2\phi_1/3$, the spatial vector component $A_x^2=\psi_p$ does not couple directly to $\tilde{A}_t$, but is affected by the background. As a result, the formation of the vector condensate becomes unfavorable compared with the growing $S_+$ scalar order.

For weak backreaction ($2b^2=0.1$), the vector condensate survives in a narrower region and coexists with the unbalanced $S_1+iS_2$ order ($S_+ +S_-$ order) in the low-temperature, low-$\mu_b$ region. However, under strong backreaction ($2b^2=1$), the $S_+$ critical temperature surpasses that of the p-wave order across the entire chemical potential domain, even at $\mu_b = 0$. The p-wave order disappears beyond a small $\mu_b$, leaving the balanced $S_+$ phase ($S_1=S_2$) as the only stable color-superconducting state throughout the high-density regime.

Across all three phase diagrams, increasing $\mu_b$ enhances the critical temperature of the $S_+$ order. This suggests that dense QCD matter may favor a $S_1+iS_2$ color-octet condensate at high $\mu_b$.

\section{Conclusions and Discussions}\label{sec5}
We have constructed a holographic model with a scalar octet coupled to both $SU(3)_c$ and $U(1)_b$ gauge fields in asymptotically AdS$_4$. The dual color-octet operator provides a complementary non-Abelian channel to the QCD diquark CSC, while the 2SC and CFL phases require a scalar in the $\bar{\mathbf 3}$ or $\mathbf 6$ representation and are left for future work. At $\mu_b=0$, the model reproduces the same phase structure in the $SU(2)$ s+p model~\cite{Nie:2013sda,Nie:2014qma,Nie:2015zia}. At finite $\mu_b$, the s-wave order generalizes to an $S_1+iS_2$ structure, and the imbalance between its components becomes essential, particularly when coexisting with p-wave order. In the probe limit, we derived analytic expressions for the critical temperatures: the p-wave critical temperature remains constant, while the $S_1+iS_2$ critical temperature increases linearly as $T_{c}^{S_+}=T_c^0(1+2\mu_{b}/(3\mu_3))$. Including backreaction deforms these analytic laws: the p-wave critical temperature decreases and vanishes at large $\mu_b$, while the $S_1+iS_2$ critical temperature still rises with increasing $\mu_b$.

The resulting $\mu_b$–$T$ phase diagrams reveal a rich competition between these orders, with the $S_1+i S_2$ critical temperature increasing with baryon chemical potential. In the high-$\mu_b$, the balanced $S_+$ phase is favored, while the $S_-$ and p-wave orders appear at low temperature and low $\mu_b$. This provides a qualitative holographic guide for color-octet condensation in dense QCD matter and may be relevant for neutron-star interiors. Future extensions include a $\bar{\mathbf 3}$/$\mathbf 6$ realization of the 2SC/CFL phases, embedding into five-dimensional holographic QCD~\cite{Chen:2022goa, Cai:2022omk, He:2023ado, Rougemont:2023gfz, Hippert:2023bel, Cai:2024eqa, Jokela:2024xgz, Shen:2025zkj}, and studying topological defect excitations.
	
\section*{Acknowledgements}
ZYN thanks Professor Yong-Chang Huang and Professor Song He for useful suggestions and discussions. This work was supported by the National Natural Science Foundation of China (Grant Nos. 12575054, 12175105, and 11965013). ZYN is partially supported by the Yunnan High-level Talent Training Support Plan Young \& Elite Talents Project (Grant No. YNWR-QNBJ-2018-181). 
\bibliographystyle{JHEP}
\bibliography{reference}
\end{document}